MOLAM: A Mobile Multimodal Learning Analytics Conceptual Framework to Support Student Self-Regulated Learning

Preprint of book chapter accepted for publication in P. Prinsloo, S. Slade, & M. Khalil (Eds.), Learning analytics in open and distributed learning: Potentials and challenges.

First submitted on: 19.01.2020
Revisions submitted on: 16.06.2020, 05.10.2020
Accepted on: 06.10.2020

Please cite as:
Khalil, M. (in press). MOLAM: A Mobile Multimodal Learning Analytics Conceptual Framework to Support Student Self-Regulated Learning. In P. Prinsloo, S. Slade, & M. Khalil (Eds.), Learning analytics in open and distributed learni ng: Potentials and challenges. SpringerOpen.

Correspondence questions and inquires about this preprint may be addressed to:

Dr. Mohammad Khalil
Centre for the Science of Learning & Technology (SLATE), University of Bergen

mohammad.khalil@uib.no

# MOLAM: A Mobile Multimodal Learning Analytics Conceptual Framework to Support Student Self-Regulated Learning

*Mohammad Khalil*
*University of Bergen*

**Introduction**

Online distance learning is highly learner-centred, requiring different skills and competences from learners, as well as alternative approaches for instructional design, student support, and provision of resources. Learner autonomy and self-regulated learning (SRL) in online learning settings are considered key success factors that predict student performance (Zhao, Chen, & Panda, 2014; Broadbent, 2017; Papamitsiou & Economides, 2019) SRL comprises processes of planning, monitoring, action and reflection (Zimmerman, 1990), and typically focuses on three key features of learners: (1) use of SRL strategies, (2) responsiveness to self-oriented feedback about learning effectiveness, and (3) motivational processes. SRL has been identified as having a direct correlation with students' success (Zimmerman, 1990), including improvements in grades and the development of relevant skills and strategies. Such skills and strategies are needed to become a successful lifelong learner.

Earlier research suggests that learners may struggle in online, open and mobile learning environments when they do not use critical SRL strategies (Wong, Khalil, Baars, de Koning, & Paas, 2019; Wong et al., 2019b). Formal, non-formal, and informal online learning settings are constantly evolving and are predominantly accessible through a range of technologies, including those provided by educational institutions (e.g., learning management systems and laptops) and students' own mobile devices (e.g., smartphones). This suggests that students need to be able to navigate effectively across a range of environments *in combination with* the use of institutionally provided and private mobile devices to succeed in their learning. It is argued that the effective use of SRL would be beneficial in these contexts, although this can be difficult for both learners and educators, particularly when students are learning online and/or independently (Lodge, Panadero, Broadbent, & de Barba, 2018).

Studies have shown that educators are challenged in helping students to develop the strategies and skills needed to regulate their learning (Lodge et al., 2018; Yen et al., 2018). Likewise, many students possess poor SRL strategies, including their ability to accurately calibrate their own learning processes (Dunlosky & Lipko, 2007). Moreover, without adequate instructional support and effective learning design, students may overestimate their understanding of learning materials which can then negatively impact on the remainder of their learning (Gyllen, Stahovich, Mayer, Darvishzadeh & Entezari, 2019). Given that increasing numbers of students are spending significant time learning independently in online flexible learning settings, "there is a growing need for understanding [i.e., measuring] and intervening [i.e., supporting] in these environments towards the development of SRL" (Lodge et al., 2018, p. 1). All this suggests that additional learner support is required for SRL development across multifaceted online learning contexts.

This chapter introduces a **Mobile Multimodal Learning Analytics** approach (MOLAM). I argue that the development of student SRL would benefit from the adoption of this approach, and that its use would allow continuous measurement and provision of in-time support of student SRL in online learning contexts.

**Background**

Although considerable theoretical and conceptual progress has been made with respect to student self-regulation in learning, there has been "little progress in developing methods to make the primary invisible mental regulation processes [...] visible and thus measurable and ultimately interpretable" (Noroozi et al., 2019, p. 2). Existing progress largely relates to those recent developments of learning analytics approaches that have focused on measuring various aspects of student SRL, frequently based on the availability of learner digital data accessed from learning management systems. However, learners today move continuously across different learning contexts (formal, informal, and non-formal), where they extensively employ their mobile devices (often smartphones) for different purposes, including learning. This type of learning is recognized as mobile learning (m-learning).

M-learning "draws on the attributes of enhanced mobility and flexibility that are enabled by portable devices and cloud-based networks" (Palalas & Hoven, 2016, p.51). M-learning as a term has been used inconsistently and to imply different meanings; it has been criticized for focusing on examining 'things' (i.e., the use of computing devices) rather than educational problems [...] that would improve learning and achieve learning goals" (Grant, 2019, p. 362). In fact, m-learning studies rarely report (or often do not report at all) underlying pedagogical and/or theoretical frameworks (Grant, 2019). This chapter aims to fill this gap by suggesting how mobile learning analytics underpinned by the theoretical lens of SRL (Zimmerman, 1990) might be designed and used to support students' learning in online learning settings.

Definitions of m-learning tend to fall into four, often overlapping, categories. These concern: (1) its relationship to distance education and eLearning, (2) exploitation of devices and technologies, (3) mediation with technology, and (4) the nomadic nature of learners and learning (Grant, 2019). Some scholars argue that such definitions are unhelpful, and researchers should instead ground their research efforts in the following design characteristics: (1) the learner is mobile; (2) the device is mobile; (3) data services are persistent; (4) content is mobile; (5) the tutor is accessible; (6) physical and network cultures and contexts impact learning or learner; and (7) the learner is engaged (Grant, 2019, p.370).

Given the importance of supporting learners in their regulation of learning activities across multifaceted online learning environments, an improved understanding and follow-up support of their SRL activities becomes critical.

Despite a broad acceptance of the various benefits of learning analytics within open, distance, and distributed educational systems to support improved retention rates and educational practices (see e.g., Khalil & Ebner, 2016b), there are few studies on learning analytics frameworks designed for use with ubiquitous mobile devices and SRL. Earlier research has shown that mobile technologies in education can be advantageous and that mobile apps can enhance students' abilities to self-regulate their learning (Broadbent, Panadero & Fuller-Tyszkiewicz, 2020). Yet, the impact of mobile technology on student learning has often been measured by individually stated perceptions rather than the direct use of technology, with some exceptions (e.g., Tabuenca, Kalz, Drachsler, & Specht, 2015; Molenaar, Horvers, & Baker 2019). This is a serious limitation for understanding the transformative nature of learning as a continuous process, and for providing opportunities to give relevant feedback and intervene in real time (i.e., to optimize student learning by improving learner support).

The mobile learning analytics field is challenging and promising for practitioners and researchers because of the distinctive features offered by mobile devices. For example, there is a large amount of *contextual* and *temporal* learner data that cannot be obtained from web-based systems but can be mined from mobile used technologies and which, in turn, impacts on the data types collected (see e.g., Toninelli & Revilla, 2020). M-learning provides opportunities to collect localized data and information from various learning activities (Tabuenca et al., 2015), continuously taking place across formal and informal learning settings. Mobile learning analytics then is explained as "the collection, analysis and reporting of the data of mobile learners, which can be collected from the mobile interactions between learners, mobile devices and available learning materials" (Aljohani & Davis, 2012, p. 71).

The chapter presents a holistic approach known as the Mobile Multimodal Learning Analytics Approach, encompassing learning analytics, m-learning, and SRL grounded in Zimmerman's (1990) theory and Aljohani & Davis' (2012) framework. It targets three types of stakeholders, namely *learners*, *teachers* and *researchers*, and aims to provide additional insight into designing, implementing and evaluating m-learning scenarios to foster students' SRL strategies, skills and knowledge in online and open learning contexts.

**Mobile Multimodal Learning Analytics Approach**

The Mobile Multimodal Learning Analytics Approach (MOLAM; Figure 1) is informed by learning design, a "methodology for enabling teachers/designers to make more informed decisions in how they go about designing learning activities and interventions" (Conole, 2012), and learning analytics, which has shown that learning design can impact on students' learning behaviour, satisfaction and outcomes (Holmes, Nguyen, Zhang, Mavrikis, & Rienties, 2019).

MOLAM can be understood and employed through the lenses of multidisciplinary and multichannel (i.e., data originating from different sources) SRL data research approaches which provide support for fostering students' SRL across formal, non-formal, and informal online learning contexts. A learner's ability to employ SRL strategies is not static; it alters throughout the learning process (Sedryakyan, Malmberg, Verbert, Järvelä, & Kirschner, 2018). Therefore, MOLAM focuses on: (1) the examination of the actual continuous uses of SRL meta-strategies, strategies and tactics, and (2) the further support accessible through learners' mobile technologies-in-use. Empirical research suggests that there is limited existing support based on learning analytics given to students to foster their SRL in m-learning settings (Matcha, Uzir, Gašević, & Pardo, 2019). In particular, the authors conclude that the prevailing part of existing SRL support is limited to web-based student-centred learning dashboards (i.e., rich visualizations). This is seen as a critical limitation in terms of learners' access and use of such web-based tools across formal, non-formal and informal learning contexts, in which learners are constantly moving, and in which they frequently employ various technologies, including their own mobile devices.

The roots of designing MOLAM were influenced by Khalil & Ebner's Learning Analytics Lifecycle model (2015) and Park's Pedagogical Framework for Mobile Learning (2011) which both build upon Zimmerman's (1990) SRL model and Winne's (2017) grain size explanation of learning analytics for SRL. The Learning Analytics Lifecycle (Khalil & Ebner, 2015) was adapted since it refines three of the early developed learning analytics models (Chatti, Dyckhoff, Schroeder, & Thüs, 2012; Clow, 2012; and Greller & Drachsler, 2012) as well as providing a solid foundation for the purposes of employing learning analytics with SRL. Park's (2011) model for m-learning delivers a sound theoretical framework to build mobile applications for the purpose of 'learning-on-the-go'. It is based on a Transactional Distance theory that is defined by *distance* not only as a geographic separation but, more importantly, as a pedagogical concept (Moore, 1997). Transactional distance is understood as the "interplay of teachers and learners in environments that have the special characteristics of their being spatially separate from one another" (Moore, 2007, p. 91). Park's model consists of individual and social aspects of learning which fits well with Zimmerman's theoretical SRL model (1990), grounded in a socio-cognitive view of SRL that includes personal, behavioural, and critical environmental classes of influence on self-regulated behaviour (Panadero, 2017).

Further development of MOLAM is expected to benefit from ongoing deployment of new digital tools that are accessible through learners' mobile technologies. The use of new technologies could: (1) contribute to learner academic success (Broadbent, 2017; Wong et al., 2019a), and (2) provide access to a new type of multichannel behavioural learner data that might reveal to researchers and practitioners how learners employ different SRL strategies and develop relevant SRL skills and knowledge over time (i.e., a process-oriented view of SRL compared to an accepted static view of SRL).

MOLAM, shown in Figure 1, consists of four key mutually constituting parts: 1) learning settings; 2) data; 3) analytics and measurement; and 4) action-support. These are explored further below.

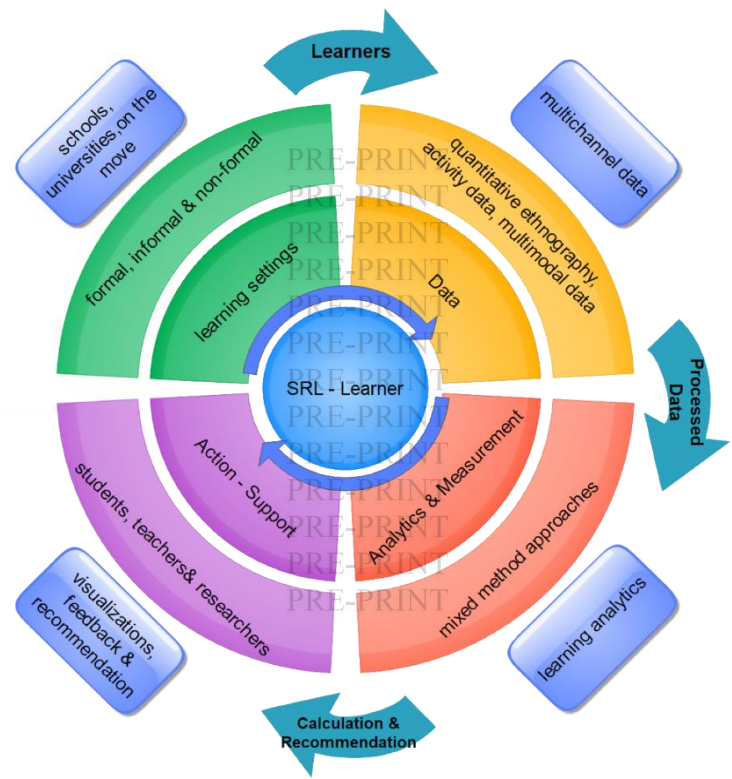

**Figure 1:** *Mobile Multimodal Learning Analytics Approach (MOLAM)*

*Learning Settings*

MOLAM targets three main learning settings: formal, informal and non-formal learning contexts:

- *Formal learning* typically implies a learning process which happens within an organised and structured context (e.g., school, college and university). It can often lead to some kind of accredited recognition (e.g., diploma, certificate).

- *Non-formal learning* usually refers to a learning process embedded in planned activities which do not include formal learning elements such as a syllabus, certification and/or

- accreditation. Examples of non-formal learning include short courses, workshops, and professional development sessions.

- *Informal learning* is defined as learning resulting from daily activities related to work, family, or leisure. It is sometimes referred to as experiential learning. It tends not to be structured in terms of learning objectives, learning time and/or learning support. Typically, it does not lead to certification. Informal learning may be also intentional (for example, when the learner takes the initiative to participate in a MOOC to learn about a certain topic). (Colardyn & Bjornavold, 2004)

Learners continuously 'move' and produce multifaceted and continuous data, including that produced through use of their own mobile technologies, as well as data related to uses of technologies for learning purposes. Smart mobile devices are extensively used in both non-formal contexts (Barbosa et al., 2016) where the educational process has a flexible curriculum, and in informal learning contexts by enthusiasts (Clough, Jones, McAndrew, & Scanlon, 2008); learners "use them in ways that correspond to the collaborative, contextual and constructivist mobile learning philosophies [...] The mobile device enthusiasts had already successfully adopted their devices and had integrated them into their daily lives*"* (p. 370)*.*

*Learner Data*

MOLAM seeks to gain insight into student learning in a mobile context along with multichannel learner data in a variety of ways, depending on the research questions and the studied context(s). In particular, it may be argued that MOLAM is grounded in a mixed-methods methodology enriched by data from multiple data sources. Sources include: 1) learner activity data, derived from use of mobile technologies, 2) quantitative ethnography (Schaffer, 2017), and 3) multimodal data collection, the analysis of which has earlier been found beneficial for the understanding of students' SRL processes (Järvelä et al., 2019). First, it is suggested that specially developed or adapted software/apps - easily accessible through mobile devices (e.g., smartphones and/or tablets) - aimed at explaining and practicing SRL in selected learning settings should be used by learners as a support tool alongside institutionalised studies. Such software needs to integrate specially designed learning analytics module/s that target different dimensions (e.g., cognitive and affective SRL processes) and phases of SRL (i.e., planning, monitoring and self-evaluation) separately or in a combination. This would provide stakeholders with process-oriented, continuous learner activity data, the analysis of which will allow mobile educational technology developers, learning designers, educators and researchers to better understand students' SRL processes over time. Second, MOLAM would benefit from the use of quantitative ethnography, an emerging methodology that integrates quantitative and qualitative methods to assess learning and human meaning-making (Schaffer, 2017). Ethnography underlines the importance of having data that is grounded; ethnographers focus on understanding "what data means to the people who are being studied" (p.110). According to Schaffer, "culture matters, because while computers can mine in a mountain of data, human beings swim in a sea of significance" (p. 20). This suggests that to be able to adequately interpret statistical SRL learner data (e.g., derived from the analysis of log data and/or more established methods, e.g., surveys), ethnographic qualitative methods of data collection should also be employed to understand the student educational culture, and the culture of their use of mobile technologies.

Third, considering that learning practices vary considerably across learning contexts, MOLAM might also involve methods for collecting multimodal learner data (e.g., spatial and proximity data, physiological measurements such as eye movement, electrodermal activity, etc) that may be accessible from mobile technologies-in-use. Researchers collecting and analysing multimodal (ethnographic) data are concerned with "accounts of cultural and social practices through prolonged fieldwork in a particular setting" (Jewitt, Bezemer, & O'Halloran, 2016, p. 118). For the analysis of complex multimodal datasets, recent computer-assisted tools, such as the Qualitative Data Analysis Software (Antoniadou, 2017) could be employed to provide valuable and practical support of complex and time-consuming qualitative research processes. For obtaining relevant multimodal process-oriented or *temporal* data, mechanisms for collecting multimodal data could be integrated through the use of students' own mobile technologies, allowing researchers to better understand the continuous nature of their SRL processes occurring across learning settings and further suggest related in-time actions aimed at improving learner support and/or teaching SRL.

Learning analytics for SRL could contribute to the development of a student-facing or teacher-facing learning dashboard - a digital instrument that can be used to visualise students' SRL processes, based on a multichannel data stream (including student log activity data from the adapted SRL software use and multimodal data), with the overall goal to facilitate the development of students' self-regulation. Making SRL processes continuously visible to learners, teachers and researchers should improve learners' ability to self-regulate their learning.

*Analytics and Measurement*

Data used in MOLAM can vary. Mixed-methods of analysis based on SRL theory and process-oriented behavioural data will provide a better understanding of the complex nature of student SRL processes (Panadero, 2017). There are many examples from quantitative learning analytics methods that can be used to support SRL using mobile generated data such as process mining and sequential pattern analysis (e.g., Wong et al., 2019a, Shabaninejad et al., 2020). Applying other data mining techniques such as decision trees and neural networks could help in classifying student types. Other types of analytics like location-aware and data moving micro clusters (Lu & Tseng, 2009) for identifying moving objects may fit well into analysing location data for the purposes of analyzing learners' behaviours for the provision of effective learning support of self-regulation (Yamada et al., 2017).

In the MOLAM proposal, I argue that the analysis of relevant learning activity data (i.e., log data from the students' use of specially adapted or developed apps aimed at fostering SRL) and other multimodal data, combined with more established and validated ethnographic methods (e.g., surveys, self-reported data, and observations) provides considerable potential to facilitate new insights into SRL (i.e., to understand), to visualise (i.e., support), and to use such data in different learning activities to inform students (Noroozi et al., 2019), teachers, and researchers. Thus, the MOLAM approach offers potential to provide improved support for not only measuring student SRL but also for fostering SRL, i.e., to optimize learning and the environments in which it occurs, thus meeting the ultimate goal of learning analytics (Siemens & Long, 2011).

*Action and Support*

The proposed approach targets three types of stakeholders; *learners*, *teachers* and *researchers* in order to optimize student learning and the environments in which it occurs, as well as to improve support for developing student SRL. Learners on the one hand, can be supported by specially developed or adapted software aimed at practicing SRL activities. The resulting mobile learning analytics could then aid teachers through the development of a teacher-facing learning dashboard that will visualise students' SRL processes, both at an individual and/or group level. Such a tool could assist teachers not only in their understanding of students' SRL processes but also in designing and practicing relevant teaching activities aiming at further fostering students' SRL in educational settings and providing adequate support. Finally, to support researchers in tracíng and interpreting students' SRL activities through a process-oriented approach, a graphical user interface to facilitate data visualisation and processing should be developed, thus offering new opportunities for researchers to explore learner data. This would contribute to a deeper understanding of the underexplored role of self-regulation in the m-learning research field and a further theoretical development of the SRL research area supported by the still emerging area of learning analytics.

**Privacy principles and MOLAM**

Several privacy and ethical concerns might emerge. According to Khalil and Ebner (2015), uses of learning analytics introduces several issues, namely: transparency, accessibility, security, ownership, policy, accuracy, and privacy. In addition to learning analytics, mobile applications are personal devices on which users store private information. The processing of personal data through mobile apps can pose significant risk to users' security and privacy. For instance, breaching one's privacy can be caused by a variety of sensors held in smart mobile devices. Examples of this are the use of location data (i.e., GPS and GLONASS), accelerometer data, microphone, camera, and Wifi, including educational apps, which create a cloud of unexpected privacy impacts.

In the context of MOLAM, privacy, confidentiality, and anonymity remain paramount. Learning analytics may reveal personal information and attitudes, as well as learner activities, which could lead to the identification of individuals to unwanted stakeholders (Khalil & Ebner, 2016a). It should be stressed then that developing mobile applications using MOLAM (and other approaches) should follow national and international frameworks such as the General Data Protection Regulation (GDPR) in the Euro zone, and the Family Educational Rights and Privacy Act (FERPA) and the Student Privacy Compass in the US. Empirical data, whether generated by prototypes or final products of learning analytics modules, should stipulate rules of transparency and consent, and ensure data protection by design and by default, as well as security of personal data processing (Castelluccia et al., 2017). Learner consent in educational contexts like educational mobile apps that fall under MOLAM should follow the general GDPR framework or its derivatives such as the blueprint by Muravyeva et al. (2020), and the JISC framework (Sclater, 2017).

**Conclusions**

SRL refers to how learners steer their own learning (Wong et al., 2019a). Fostering SRL for students in distance education can be particularly challenging (Andrade & Bunker, 2011). Increasing demands to understand and support learners' SRL activities in open, distance, and distributed systems (including m-learning environments) require further employment of innovative teaching and learning practices.

This chapter introduces MOLAM, a model aimed at guiding learners, teachers and researchers wanting to develop, successfully employ and/or evaluate learning analytics approaches for mobile learning activities for the purposes of measuring and fostering student SRL in diverse online learning environments. MOLAM is especially valuable for continuous measurement and interventions, thus fostering students' transferable SRL skills, strategies and knowledge across formal, informal and non-formal online learning settings. This would benefit not only learners' academic success but also their development as successful lifelong learners.

In conclusion, it is suggested that:

1. There is a need to measure and support SRL in digital and distance learning to enable learners to take greater control over their own learning, underpinned by sound theoretical models and frameworks. The field of mobile learning analytics has the potential to support this given its powerful multidisciplinarity nature. It is anticipated that learning analytics methods will deal with multimodal multichannel data from various dimensions associated with SRL.

2. Although the collection, use and analysis of multimodal multichannel data continues to evolve within learning analytics, researchers should address challenges resulting from instrumentation errors, reliability of measures, experimental designs, and inferences about process data (Azevedo & Gašević, 2019).

3. The application of mobile multimodal learning analytics should be performed with careful integration of relevant support mechanisms and frameworks to protect student privacy and ensure their agency in online learning settings.